\documentstyle[12pt,amssymb,epsfig]{article}
\author {Ernesto S. Loscar$^{a}$ and Ezequiel V. Albano$^{a}$\\
$^a${\it Instituto de Investigaciones Fisicoqu\'{\i}micas
Te\'{o}ricas y Aplicadas}\\{\it (INIFTA), UNLP, CONICET, 
Suc.4, CC16,}\\{\it
1900 La Plata, Argentina}}

\title{Numerical study of a first-order irreversible phase transition
in a $CO+NO$ catalyzed reaction model.}
\begin{document}
\maketitle


\begin{abstract}
The first-order irreversible phase transitions (IPT) of the Yaldran-Khan
model (Yaldran-Khan, J. Catal. {\bf131}, 369, 1991) for the $CO+NO$ reaction 
is studied using the constant coverage (CC) ensemble and performing 
epidemic simulations.
The CC method allows the study of hysteretic effects close to coexistence
as well as the location of both the upper spinodal point and the coexistence
point. Epidemic studies show that at coexistence the number of active sites
decreases according to a (short-time) power law followed by a (long-time)
exponential decay. It is concluded that first-order IPT's 
share many characteristic of their reversible counterparts, such as the 
development of short ranged correlations, hysteretic effects, metastabilities,
 etc.
\end{abstract}

\section{Introduction}

The study of critical phenomena occurring in adsorbed overlayers is a topic
that has attracted increasing attention \cite{binder}. In fact, the
understanding of the behavior of atoms and molecules absorbed on different
surfaces is essential for many branches of science (chemistry, physics,
biology, etc). Within this wide context, surface phenomena occurring under 
non-equilibrium conditions are far from being understood. Therefore, they have
become the center of great attention \cite{dicklib}. In particular,
irreversible catalytic reactions exhibit a very rich and complex behavior 
that includes oscillations, bifurcations, chaos metastability,
irreversible phase transitions (IPT's), etc. \cite{n3,ykmodel,revhcr,revlib,zgb,libaleman,ehashi}.
  
The aim of this work is to perform an extensive numerical study of the
first-order IPT characteristic of a model for the $NO+CO$ reaction early
proposed by Yaldran
and Khan (YK) \cite{ykmodel}. IPT's in reactive systems take place between an
active regime with sustained outcome of the reaction product from the
catalytic surface and an absorbing (or poisoned) state where the
catalyst becomes fully covered by one or more types of reactants. Since
the systems cannot escape from the absorbing state the transitions are
irreversible \cite{dicklib,revhcr,revlib}.
The study of IPT's has gained increasing attention in the field of
nonequilibrium statistical physics since the pioneering work of Ziff, 
Gulari and Barshad (ZGB) that introduced a simple lattice model for the
catalytic oxidation of $CO$ \cite{zgb}. Particularly interesting are
first-order IPT's that can be characterized by an abrupt change in the
density of reactants and in the rate of production due to a 
tiny change in the (external) control parameter, which is usually
the pressure. In fact, evidence of such kind of transition has been
reported in  catalyzed reactions \cite{libaleman,ehashi}. The
occurrence of hysteretic effects around a first-order transition point
has also been observed \cite{histerjcp}. Very recently, we have
performed a detailed study of the first-order IPT of the ZGB model 
presenting conclusive evidence of hysteretic effects \cite{ezer}.
As anticipated above, the aim of this work is to characterize
the first order IPT of the YK model \cite{ykmodel} for the $CO+NO$ reactions
by means of Monte Carlo simulations with particular emphasis on
the occurrence of hysteretic effects and by performing useful comparisons
with our previous study of the ZGB model \cite{ezer}.

The manuscript is organized as follows: in Section II we describe the
YK model and three different simulation methods used in the work, namely
the standard ensemble, the constant coverage ensemble and the epidemic
approach. In Section III we present and discuss the obtained results,
while our conclusions are stated in Section IV.

\section{The model and Monte Carlo simulation methods.}

\subsection{The YK model for the $CO+NO$ reaction.}

Yaldran and Khan \cite{ykmodel} have proposed a lattice gas model for the
catalytic reaction of $CO+NO$ based on the Langmuir-Hinshelwood mechanism.
The reaction steps are as follows:

\begin{equation}
NO(g) + 2 S \rightarrow N(a) + O(a)
\label{noad}
\end{equation}
\begin{equation}
CO(g) + S \rightarrow CO(a)
\label{coad}
\end{equation}
\begin{equation}
CO(a) + O(a) \rightarrow CO_{2}(g) + 2 S
\label{reco}
\end{equation}
\begin{equation}
N(a) + N(a) \rightarrow  N_{2}(g) + 2 S
\label{ren}
\end{equation}
where $S$ represents an unoccupied site on the catalyst surface, $2S$
represents a nearest neighbor (NN) pair of such sites, $(g)$ indicates
a molecule in the gas phase and $(a)$ indicates an species adsorbed on the
catalyst. The reactions given by equations (\ref{reco}) and (\ref{ren})
are assumed to be instantaneous (infinity reaction rate limit) while the
limiting steps are the adsorption events given by equations (\ref{noad}) and
(\ref{coad}). The YK model is similar to the ZGB model for the
$CO + O_{2}$ reaction, except that the $O_{2}$ is replaced by $NO$, and
NN  $N$ atoms, as well as NN $CO-O$ pairs, react. For further details
on the YK model see \cite{ykmodel,bziff,menga,mengb,adick}.

\subsection{Monte Carlo simulation method using the standard ensemble}

Monte Carlo simulations are performed on the hexagonal (triangular) lattice
of side L, assuming periodic boundary conditions. In the standard ensemble
the procedure is as follow: let $P_{NO}$ and $P_{CO}$ be the relative
impingement rates for $NO$ and $CO$, respectively, which are taken to
be proportional to their partial pressures in the gas phase. Taking
$P_{CO}\,+P_{NO}\, =\,1$, such normalization implies that the YK model
has a single parameter that is usually taken to be $P_{CO}$. $CO$ and
$NO$ adsorption events are selected at random with probabilities
$P_{CO}$ and $1-P_{CO}$, respectively. Subsequently, an empty site
of the lattice is also selected at random. If the selected species is
$CO$, the adsorption on the empty site occurs according to
equation (\ref{coad}). If the selected molecule is $NO$, a NN site of
the previously selected one is also chosen at random, and if such site
is empty the adsorption event takes place according to equation
(\ref{noad}). Of course, if the NN chosen site is occupied the adsorption
trial is rejected. After each successful adsorption event all NN sites
of the adsorbed species are checked at random for the occurrence of
the reaction events described by equations (\ref{reco}) and (\ref{ren}).

The Monte Carlo time step (MCS) involves ${L}^{2}$ adsorption attempts,
so that every site of the lattice is selected once, on average.
Simulations are started with empty lattices. The first $10^4$ MCS are
disregarded to allow the system to reach the stationary regime and
subsequently, averages are taken over $4$ x $10^4$ MCS.
During the simulations, the coverages with $CO$, $O$ and $N$ 
($\theta_{CO}$, $\theta_{O}$ and $\theta_{N}$, respectively) as well
as the rate of production of $CO_{2}$ and $N_{2}$
($R_{CO_{2}}$, $R_{N_{2}}$, respectively) are measured.
The phase diagram of the YK model is similar to that of the ZGB model 
\cite{zgb}, in the sense
that both second- and first- order IPT's are observed. However, in
contrast to the ZGB model where the absorbing (poisoned) states are unique,
in the case of the YK such states are mixtures of
$O(a)+N(a)$ and $CO(a)+N(a)$ as follows from the observation of the 
left and right sides of the phase diagram, respectively (figure 1(a)).

The IPT observed close to $P^{1}_{CO}=0.184(1)$ is continuous and therefore
of second-order (see figure 1).
Our estimation of $P^{1}_{CO}$ is in agreement with previous calculations,
namely $P^{1}_{CO}=0.185(5)$ (reference \cite{bziff}) and $P^{1}_{CO}=0.185(2)$
(reference\cite{ykmodel}).

More interesting, an abrupt first-order IPT is also observed close to  
$P^{2}_{CO}=0.3545(5)$ (figure 1(a) and (b)), in agreement with previous simulations  
(references \cite{ykmodel,bziff}). It should be noticed that a reactive
window for the YK model is observed on the hexagonal lattice (see figure 1)
while such window is absent on the square lattice 
\cite{ykmodel,bziff}, pointing out the relevance of
the coordination number on the reactivity.

\begin{figure}
\centerline{{\epsfxsize=2.5in \epsfysize=1.57in \epsffile{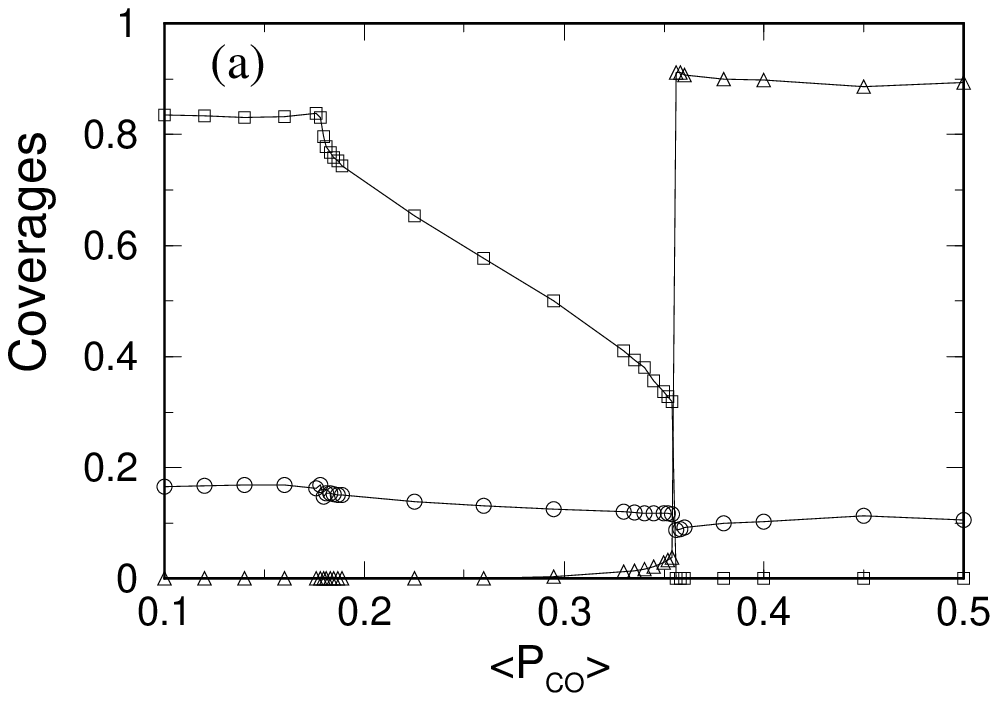}}}
\centerline{{\epsfxsize=2.5in \epsfysize=1.57in \epsffile{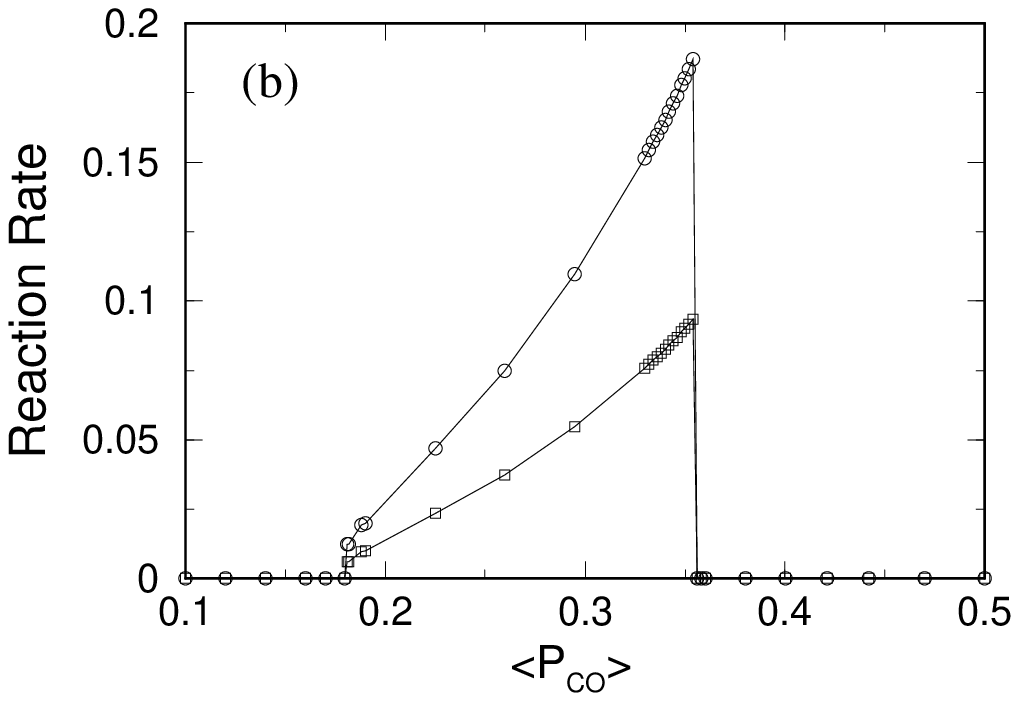}}}
\caption{Phase diagram of the YK model on the hexagonal lattice of size
L=128 LU. (a) Plots of ${\theta}_{CO}(\bigtriangleup)$, 
${\theta}_{O}(\Box)$ and ${\theta}_{N}(\bigcirc)$ versus $<P_{CO}>$. 
(b) Plots of $R_{N_{2}}(\Box)$ and $R_{CO_{2}}(\bigcirc)$;
measured in units of number of $N_{2}$ and $CO_{2}$ molecules removed
from the lattice per unit of area and time, respectively;
versus $<P_{CO}>$.}
\label{fig 1}
\end{figure}

\subsection{The constant coverage ensemble}

The constant coverage ensemble (CC) was early introduced by Ziff and Brosilow
\cite{zb} to study the first-order IPT of the ZGB model. The first step 
of the CC ensemble is to obtain a stationary configuration using the
standard ensemble. 
Then the system is actually switched to the CC ensemble, where the density 
$\theta_{CO}$ is now kept as constant as possible. For this purpose, if
$\theta_{CO}$ is below the established value, $CO$ is adsorbed on a randomly
selected site. Otherwise, if $\theta_{CO}$ is greater than the desired value, 
$NO$ adsorption attempts on randomly selected sites are performed. 
Now the effective $CO$-pressure ($<\,P_{CO}\,>$) is given by the ratio
of $CO$-adsorption attempts to the total number of adsorption
attempts.

It is worth mentioning that using the CC method as proposed by
Ziff and Brosilow \cite{zb}, the actual coverage with $CO$ is 
not strictly constant but it is affected by fluctuations of the
order of $\approx 1/L^{2}$. Very recently \cite{tania}, a true 
constant coverage ensemble, where the number of particles remains
strictly constant, has been proposed and used to study the
conserved contact process. This new method seems to be very useful
for the study of single particle systems, such as the contact 
process, branching annihilating walkers, etc., although
their implementation for complex multiparticle systems, such
as the YK model, does not appear to be straightforward.  

In simulations, $<\,P_{CO}\,>$ is averaged over $\tau_{M}$ time steps.
Subsequently, $\theta_{CO}$ is increased stepwise and measurements of 
$<\,P_{CO}\,>$ 
are performed after some waiting time $\tau_{W}$ to allow the relaxation
of the system. In this way the growing branch (GB) of the CC loop can be 
obtained. Then, after reaching a large value of $\theta_{CO}$ 
($\theta_{CO}\,\approx\,0.88$ in this work in order to prevent the
system from reaching an absorbing state as shown in figure 1), $\theta_{CO}$ 
is decreased stepwise. Using this procedure the decreasing branch (DB)
of the CC loop can be recorded. It should be noticed that in the CC ensemble
$\theta_{CO}$ assumes the role of the control parameter. Further details 
on the CC method can be found in references \cite{revlib,ezer}.      

\subsection{Epidemic simulations close to coexistence.}

Another powerful approach to the study of IPT's is to perform the so-called
epidemic analysis (EA) \cite{ezer12,ezer13}. In EA the simulations 
start from a configuration very close to the absorbing state and 
subsequently the time evolution of the system is followed using the 
standard ensemble (SE). 
In order to generate the initial configuration, a natural absorbing 
state has to be achieved first using the SE and following the 
dynamics of the system. This procedure 
assures the development of the characteristic correlations among the reactants.
Taking the obtained absorbing state, a small patch of empty sites is created 
close to the center of the sample. Subsequently, during the time evolution
of the system the average number of empty sites ($N(t)$) and 
the survival probability $P(t)$
of the active state are recorded. Each single EA stops when the 
system is trapped in the absorbing state 
so that $N(t)=0$. In order to obtain reliable results, the quantities 
of interest have to be averaged over a large number of independent 
EA ($\sim$$3$x$10^9$ runs in the present work).

Performing EA close the second-order IPT the scaling ansatz 
$N(t)\,\propto\,t^{\eta}$, where $\eta$ is an exponent,
has been proposed to hold at criticality \cite{ezer12,ezer13}.
This observation is in agreement with well established ideas 
developed by studying equilibrium (reversible) phase transition: 
scale invariance reflects the existence of a divergency in the 
correlation length. However, close to 
first-order transition it is also well known that correlations are
short ranged, preventing the observation of scale invariance. Recently 
Monetti and Albano \cite{ezer} have proposed that, at coexistence, $N(t)$
should decrease according to a short-time power law followed by a long-time
exponential decay, so that:
\begin{equation}
N(t)\propto(\frac{t}{T})^{-\eta_{eff}} \exp[-(\frac{t}{T})]
\label{anz}
\end{equation}
where $T$ sets a characteristic crossover time scale and $\eta_{eff}$ 
is an effective exponent. It should be noticed
that equation (\ref{anz}) holds for the ZGB model at coexistence \cite{ezer}.
   
\section{Results and discussion.}
       
\subsection{Study of the hysteretic effects using the CC ensemble.}

In order to study hysteretic effects CC simulations using 
lattices of different sizes and various values 
of $\tau_{W}$ have been performed. Since the measurement in  
time ($\tau_{M}$) has to be reduced in order to observe 
hysteretic effects, results are averaged over several ($10-100$) 
different loops, depending on the 
lattice size, in order to obtain acceptable statistics.

Figure 2 shows a plot of $\theta_{CO}$ vs. $<\,P_{CO}\,>$ as obtained using 
$L=32 LU$, $\tau_{W}=50$ MCS, and $\tau_{M}=50$ MCS. 
For this small lattice size the
relaxation time is quite short, so that hysteretic effects are absent. 
This result is in agreement with similar measurements performed 
applying the CC 
ensemble to the ZGB model \cite{ezer}. Note that a narrow vertical
region close to the center of the  loop for ($<\,P_{CO}\,>\,\approx\,0.35$)
can also be observed in figure 2. On the other hand, the $L$-dependent
upper spinodal point ($P_{CO}^{US}$) can also clearly be observed as shown 
in figure 2. On increasing the lattice size ($L=64 LU$, in figure 3), 
the onset of hysteretic effects can be observed 
for $\tau_{W}=75$ MCS (figure 3(a)), while 
such effects become almost negligible if the waiting time is increased 
(figure 3(b) for $\tau_{W}=400$ MCS). Furthermore, as in the case 
of figure 2, a vertical region located at
the center of the loop and slightly above $<\,P_{CO}\,>\,\approx\,0.35$ 
can be observed in figure 3(a). Also notice that such region becomes 
well defined when $\tau_{W}$ is increased (figure 3(b)).

\begin{figure}
\centerline{{\epsfysize=3.0 in \epsffile{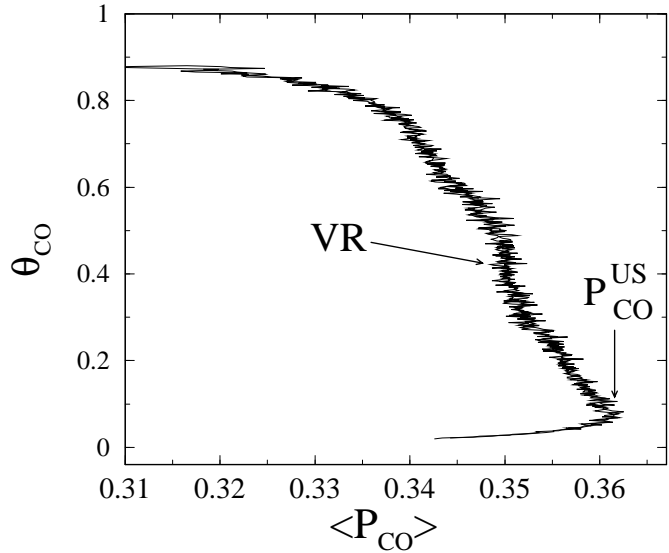}}}
\caption {Plots of $\theta_{CO}$ versus $<P_{CO}>$ obtained using the CC 
ensemble with $L=32$ LU, $\tau_{R}=50$ MCS and $\tau_{M}=50$ MCS. 
The arrows show
the narrow vertical region of the loops (VR) and the upper spinodal point 
($P^{US}_{CO}$), respectively. More details in the text.}
\label{fig 2}
\end{figure}

On increasing the lattice the hysteretic effects can be observed even 
taking larger 
values of $\tau_{W}$. In fact, for $L=128 LU$ (figure 4(a)) 
and $L=256$ (figure 4(b)), hysteresis is 
still observed for $\tau_{W}=1700$ MCS and $\tau_{W}=2700$ MCS respectively. 

\begin{figure}
\centerline{{\epsfysize=3.5 in \epsffile{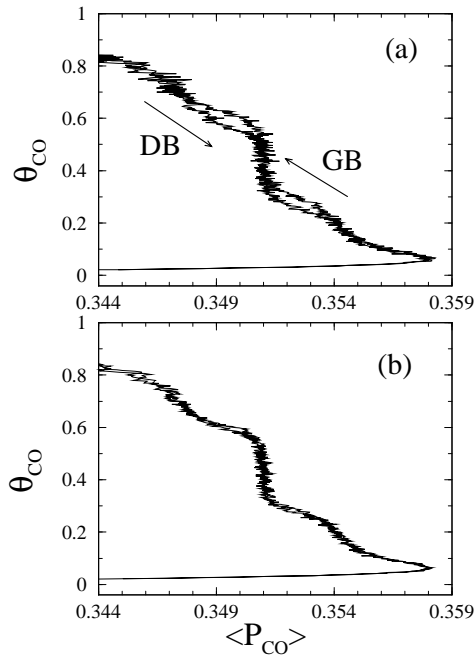}}}
\caption{ Plots of 
$\theta_{CO}$ versus $<P_{CO}>$ obtained using the 
CC ensemble with $L=64$ LU, and (a) $\tau_{W}=75$ MCS, $\tau_{M}=50$ MCS and
(b) $\tau_{W}=400$ MCS, $\tau_{M}=100$ MCS. The arrows pointing up and down, 
in figure (a), show the growing and decreasing branches (GB and DB) of the 
loop, respectively. More details in the text.}
\label{fig 3}
\end{figure}

It should be noticed that the loop becomes narrow when the lattice size is
increased, as follows from the comparison of figures (3) and (4). 
Comparing these figures it also becomes evident that while the 
location of $P^{US}_{CO}$ is shifted systematically toward lower 
values when $L$ is increased, the location of the vertical
region (close to the center of the loops) remains almost fixed very close to 
$P_{CO}=0.3515$ (see figure 4(b)).

The evaluation of a CC loop for $L=1024 LU$ requires 
huge CPU resources, so we have
restricted ourselves to the case $\tau_{W}=600$ MCS 
and $\tau_{M}=700$ MCS, since considerably larger values of 
$\tau_{M}$ became prohibitive. In this case (figure 5) hysteretic 
effects are quite evident and also, the growing and decreasing 
branches of the loops are almost vertical. However, 
the vertical region at the center of the 
loop, previously observed using smaller lattices, is no longer found.

\begin{figure}
\centerline{{\epsfysize=3.5 in \epsffile{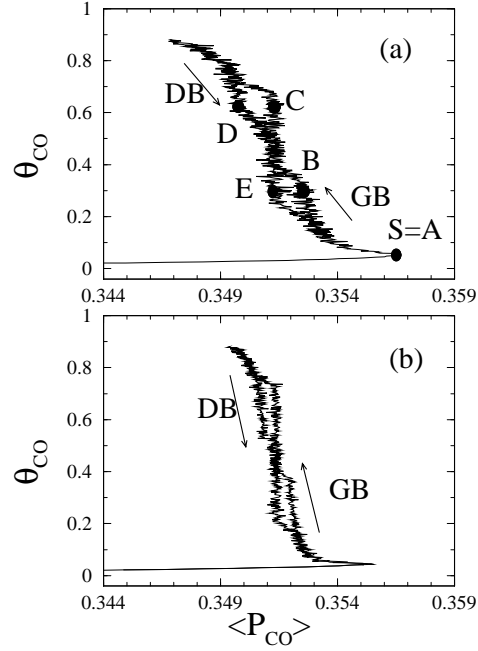}}}
\caption{ Plots of
$\theta_{CO}$ versus $<P_{CO}>$ obtained using the 
CC ensemble and taking: (a) $L=128$ LU, $\tau_{W}=1700$ MCS, 
and $\tau_{M}=300$ MCS. The points labeled A, B, C, D and E 
correspond to the coverages
used to obtain the snapshot configurations shown in figure 
6 (a)-(e), respectively. (b) $L=256$ LU and $\tau_{W}=2700$ MCS 
and $\tau_{M}=300$ MCS.}
\label{fig 4}
\end{figure}
\begin{figure}
\centerline{{\epsfysize=1.8 in \epsffile{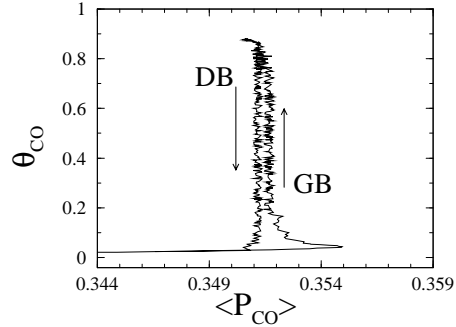}}}
\caption{Plots of 
$\theta_{CO}$ versus $<P_{CO}>$ obtained using the 
CC ensemble for (a) $L=1024$ LU, and $\tau_{W}=600$ MCS, 
and $\tau_{M}=700$ MCS.}
\label{fig 5}
\end{figure}

In order to gain insight into the behavior of the system close to coexistence
it is useful to analyze snapshot configurations, as shown in figure 6. 
Snapshots are obtained for some relevant points as indicated in figure 4(a).
Just at the upper spinodal point (figure 6(a)) one observes that the active 
phase remains homogeneous but the nucleation of few $CO$-clusters has 
already started. The biggest one (close to the lower right hand side in 
figure 6(a)) is compact (except for the presence of few $N$-atoms 
embedded into the bulk of the $CO$-cluster)
and has reached the critical nucleation size. At the upper spinodal
point one has $\theta_{CO}\,\approx\,0.045$. On increasing the $CO$ coverage 
this critical nucleus grows up into a solid and compact $CO$ cluster 
surrounded by the (homogeneous) active phase, as shown in figure 6(b). 
The coverage with $CO$ has increased up to $\theta_{CO}\,\approx\,0.28$, 
the compact $CO$ cluster does not percolate and its interface 
is essentially convex. Further increasing $\theta_{CO}$ causes the 
$CO$ cluster to percolate along one direction of the sample, 
figure 6(c). The hysteresis loop also changes for the growing 
branch (as in figure 6(b)) to the central region with 
$\theta_{CO}\,\approx\,0.63$ in figure 6(c). The percolating cluster 
is quite stable and has an interface essentially flat with a 
length of the order of $2L$. An additional 
growth of $CO$ coverage causes the $CO$ cluster to 
percolate along both directions of the lattice (figure 6(d)). 
Here the interface of the cluster is 
essentially concave and the active phase is surrounded by the massive cluster. 
Since the coverage with $CO$ is the same ($\theta_{CO}\,\approx\,0.63$) in 
both figures 6(c) and 6(d), the presence of hysteretic effects 
has to be related to the curvature 
of the interface of the $CO$ cluster. Subsequently, on 
decreasing $\theta_{CO}$ 
one observes the formation of a $CO$ cluster that percolates along only one
direction of the sample (figure 6(e)). Such kind of clusters are observed 
along the vertical region and are characterized by an almost flat interface 
with an infinite effective curvature radius. Notice that despite the fact
that the $CO$ coverage is the same in both figures 
6(b) and 6(e) ($\theta_{CO}\,\approx\,0.28$), the curvature of 
the interface of the $CO$ cluster is quite different 
and hysteretic effects are observed (see figure 4(a)).

\begin{figure}
\centerline{{\epsfysize=2.0 in \epsffile{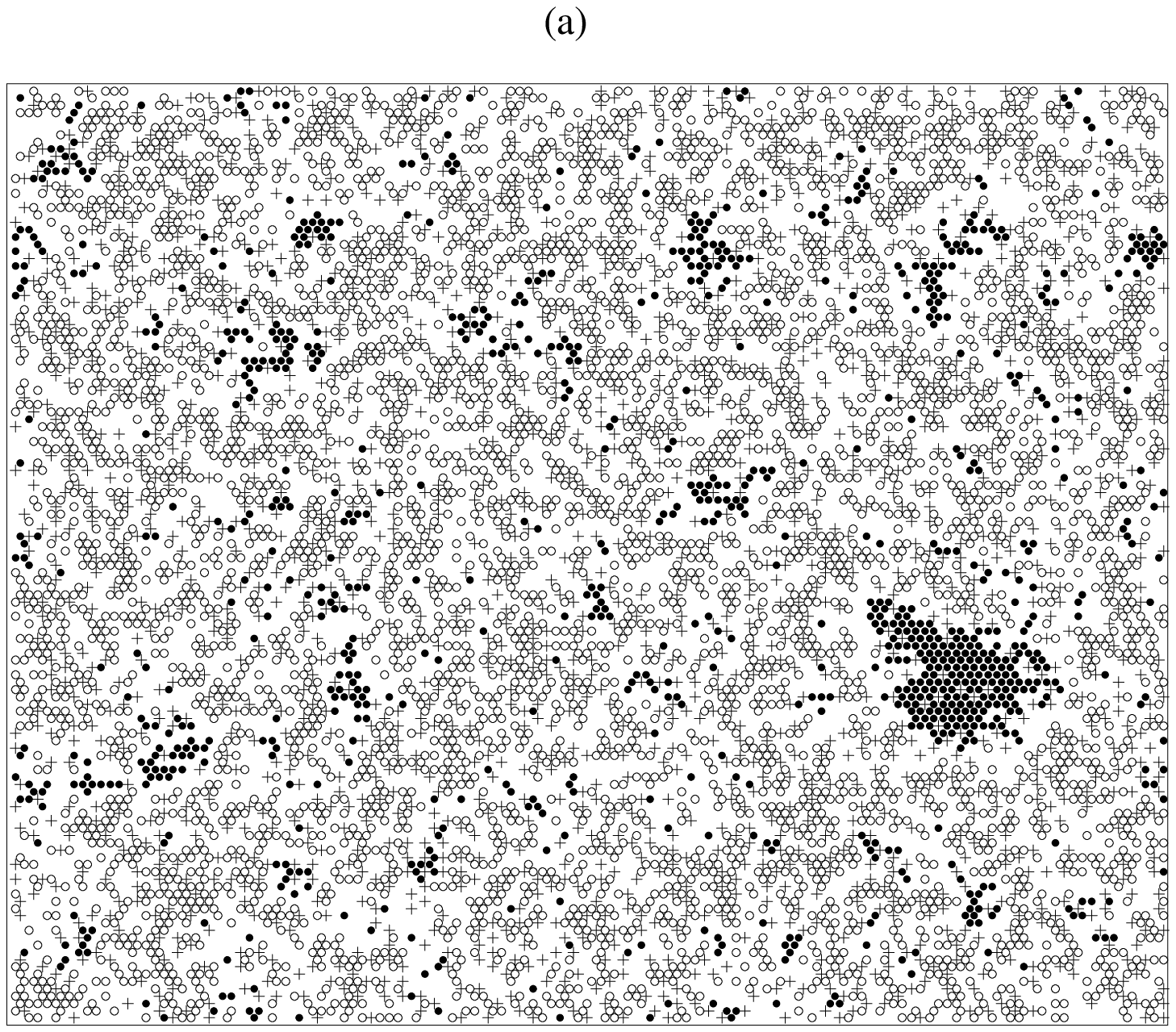} \epsfysize=2.0 in \epsffile{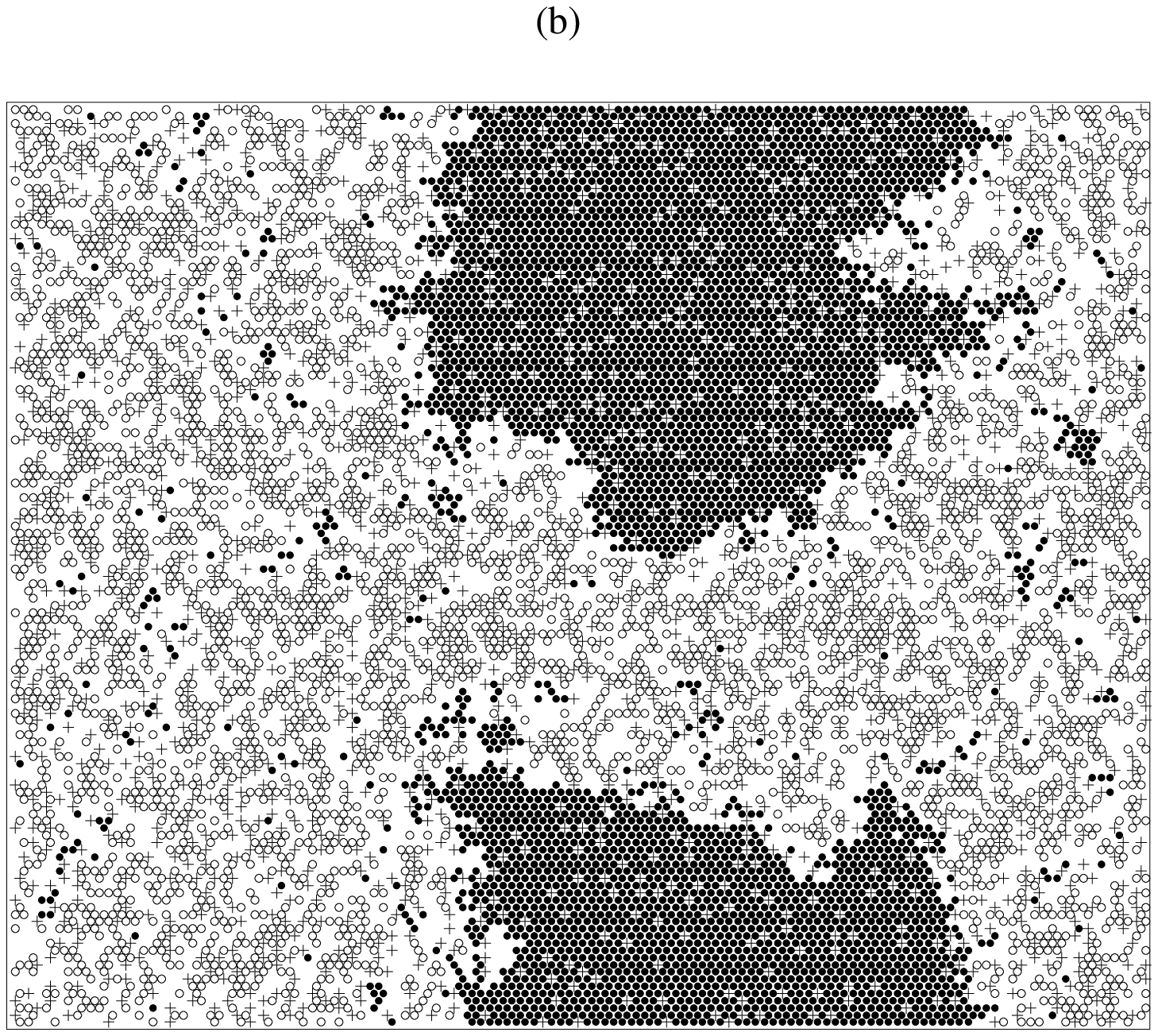}}}
\centerline{{\epsfysize=2.0 in \epsffile{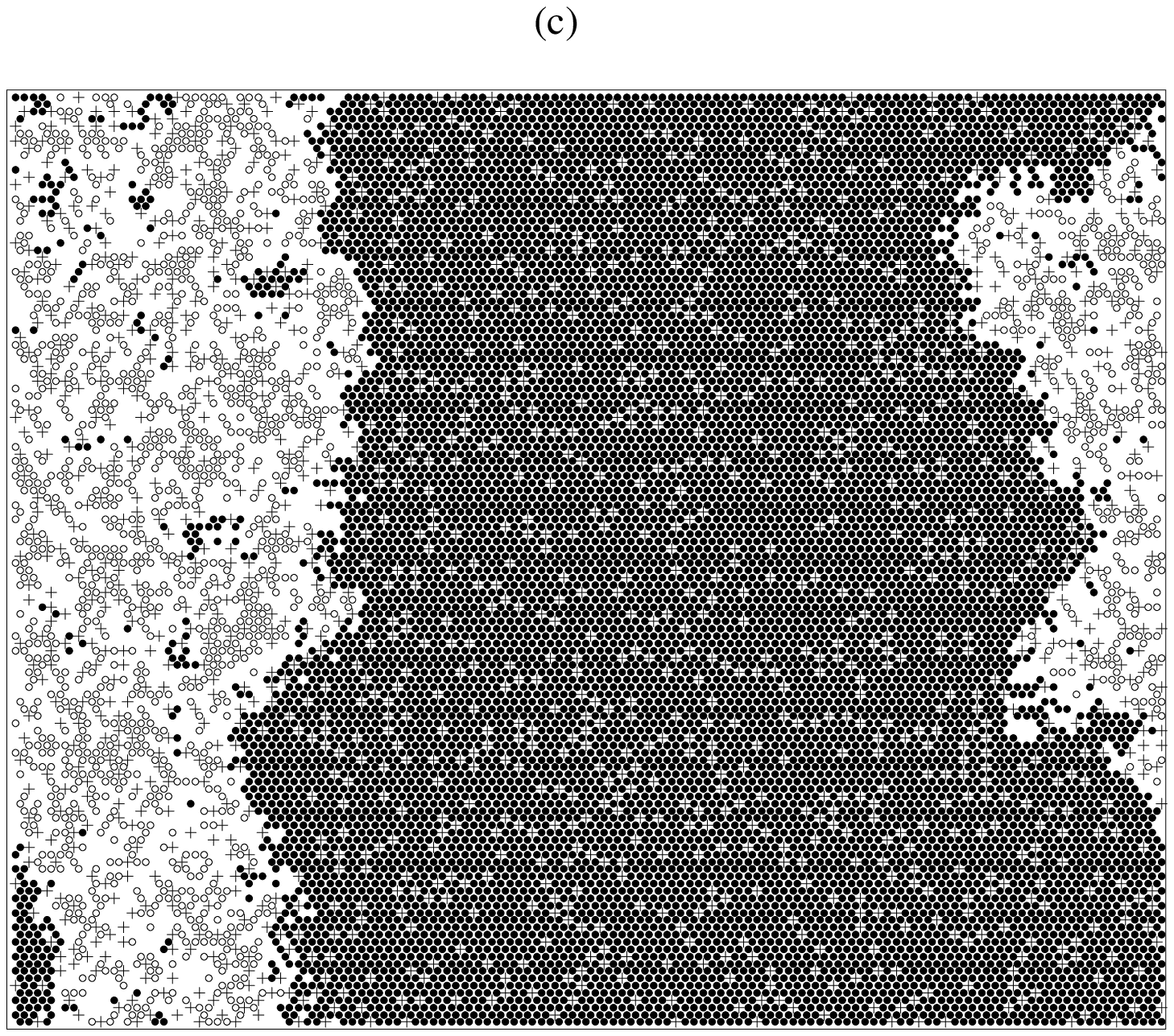} \epsfysize=2.0 in \epsffile{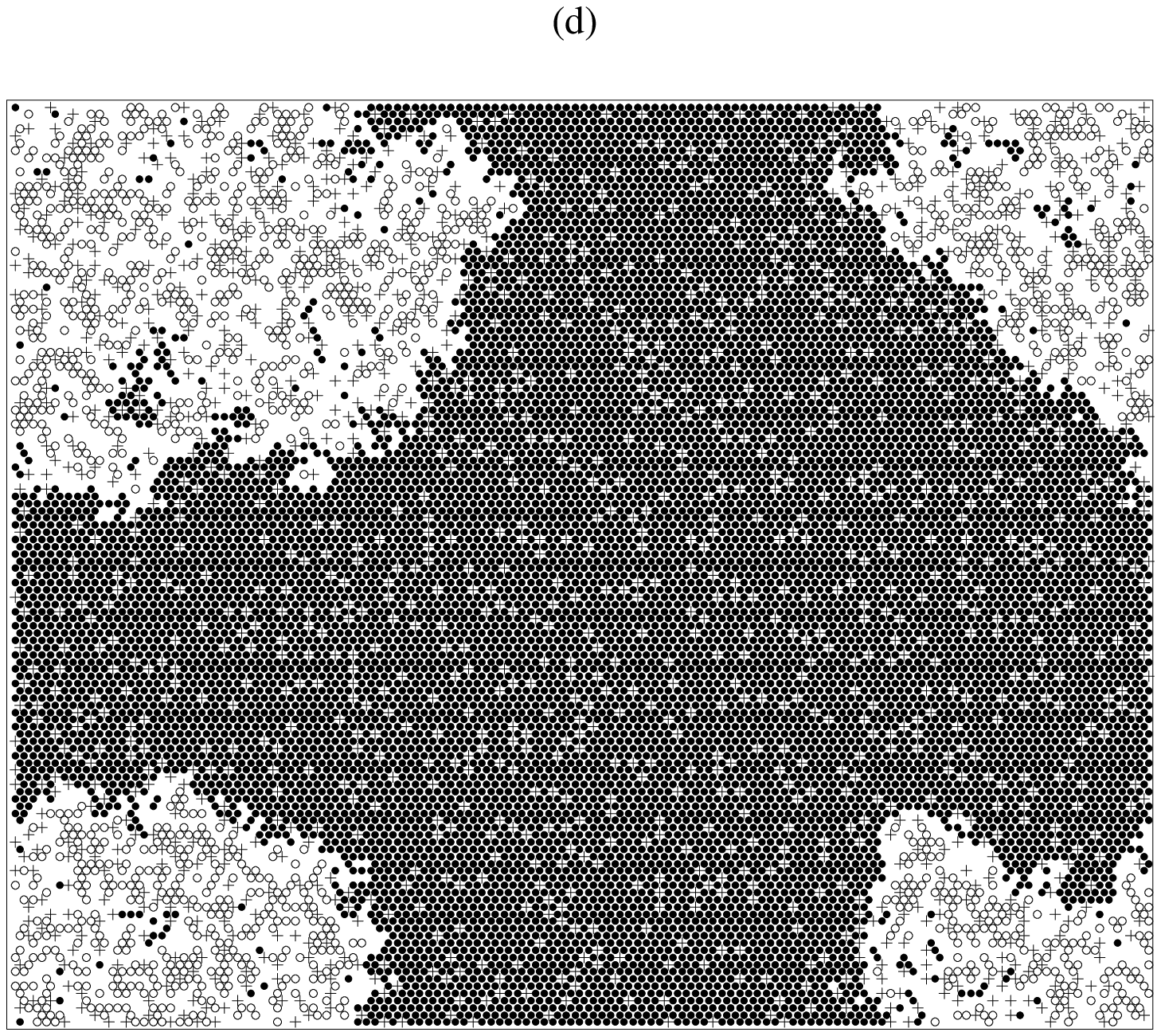}}}
\centerline{{\epsfysize=2.0 in \epsffile{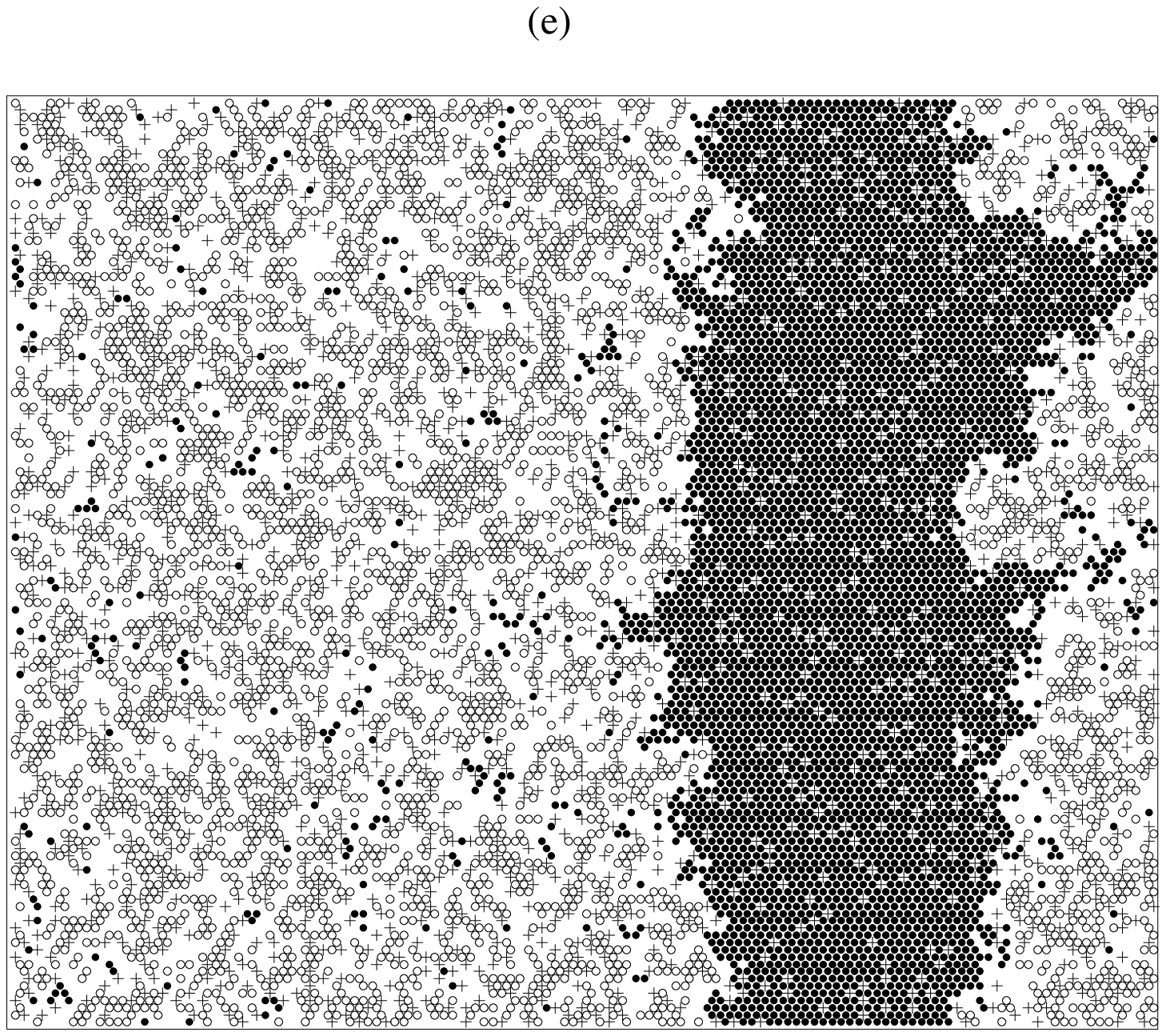}}}


\caption{Typical snapshot configurations obtained using the 
CC ensemble close to coexistence for $L=128$ LU and the points 
A, B, C, D and E shown in figure 4(a).
(a) Snapshot corresponding to the upper spinodal point S$\equiv$A, (b) and 
(d) snapshots corresponding to the points B and D respectively. (c) and (e) 
$CO$-percolating clusters obtained in the central region 
(points C and E respectively). Here $\circ$ denotes $O$(ad), 
$+$denotes $N$(ad) and $\bullet$ denotes $CO$(ad).}
\label{fig 6}
\end{figure}

Summing up, the sequence of snapshots shown in figure 6, as well as additional
figures not shown here for the sake of space, allows us to infer the 
following behavior of the interface of the $CO$ cluster along the hysteresis
loop: between the spinodal point and the growing branch the curvature 
radius of the interface is positive and finite. Along the vertical 
region the radius of curvature is infinite, while within the 
decreasing branch the radius of curvature is negative and finite. 
Almost horizontal regions observed in the loops 
correspond to regions where the curvature radius  crosses 
over between two different behaviors.

In order to quantitatively analyze the behavior of $P_{CO}$ corresponding to 
the different branches and the vertical region, it is assumed that the growing
(decreasing) branch starts at the point where both curves merge (split out).
Figure 7 shows the dependence of the location of the growing branch and the 
decreasing branch ($P^{GB}_{CO}$ and $P^{DB}_{CO}$, respectively) 
on the inverse of the lattice size. The $L-$dependence of $P_{CO}$ at 
the vertical region ($P^{VR}_{CO}$)
has also been included for the sake of comparison.

\begin{figure}
\centerline{{\epsfysize=3.0 in \epsffile{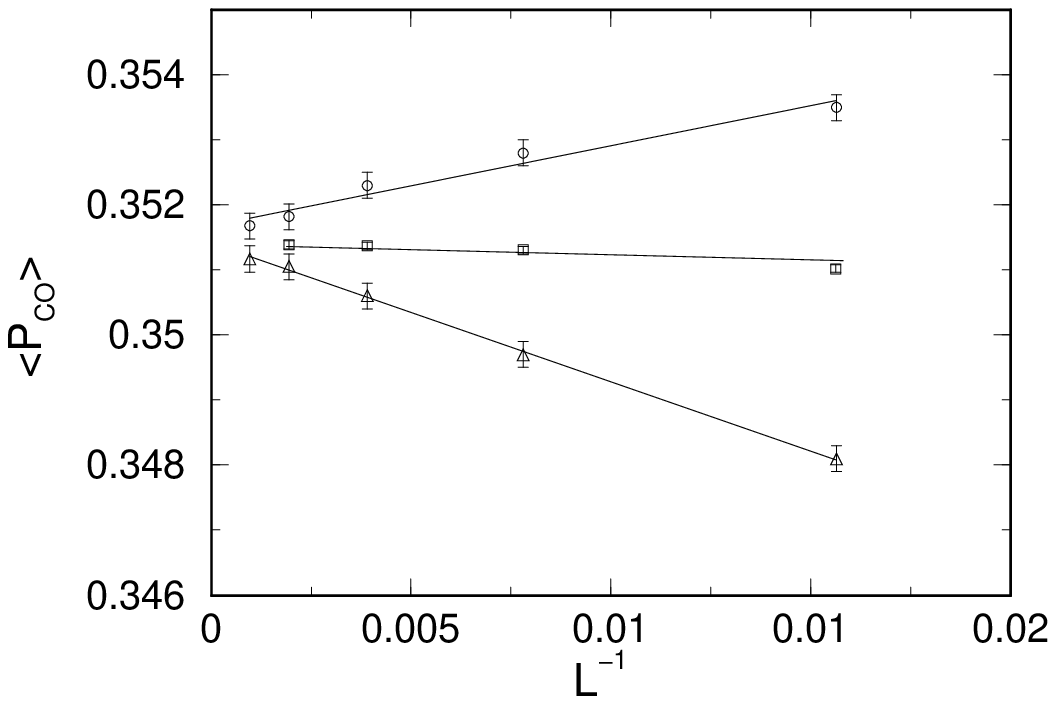}}}
\caption{Plots of
 $<P_{CO}>$ versus $L^{-1}$ measured in the growing branch ($\bigcirc$), 
decreasing branch ($\triangle$), and the vertical region
($\square$). 
The straight lines correspond to the best fits of the 
data that extrapolate to  $L\rightarrow\infty$. $L$ is measured in
LU.}
\end{figure}

As already discussed the location of all relevant points, namely 
$P^{X}_{CO}$ with $X=GB,DB$ and $VR$ depend on the curvature radius ($s$)
of interface of the massive $CO$ cluster. Such dependence can be written as 
follows:
\begin{equation}
P^{X}_{CO}=P^{X}_{CO}(L\rightarrow\infty)+F^{X}(s) {,}
\end{equation}
where $P^{X}_{CO}(L\rightarrow\infty)$ is the location of the point under 
consideration after proper extrapolation to the thermodynamic limit and 
$F(s)$ is an $s$-dependent function. For the vertical region one has 
$s\rightarrow\infty$ and $P^{VR}_{CO}$ is almost independent of $L$, 
so $F^{VR}(\infty)\rightarrow\,0$, as shown in figure 7. In contrast, for
the DB and the GB , $s$ is finite and of the order of 
$-1/L$ and $1/L$, respectively. 
So, one has $F^{DB}(s)\,\approx\,-A/L$ while $F^{GB}(s)\,\approx\,B/L$. All 
these arguments can be confirmed by the results shown in figure 7 
and the extrapolated points are:
            
\begin{center}
$P^{GB}_{CO}(L\rightarrow\infty)=0.3514(3){,}$

$P^{DB}_{CO}(L\rightarrow\infty)=0.3517(3){,}$

$P^{VR}_{CO}(L\rightarrow\infty)=0.35145(5){,}$
\end{center}
respectively. Also, $A\,\approx\,0.215(5)$ and 
$B,\approx\ 0.12(2)$ are obtained.

The observed behavior allows us to identify 
$P^{VR}_{CO}(L\rightarrow\infty)$ as the coexistence 
point $P^{Coex}_{CO}\,\cong\,0.35145(5)$
in excellent agreement with the value $P_{CO}=0.35140(1)$ reported by
Brosilow and Ziff (\cite{bziff}).

This result is in contrast with measurements performed with the ZGB model.
In fact, for the ZGB systems the vertical region is not observed while the
locations of the growing and decreasing branches are almost independent of the
lattice size \cite{ezer}. Consequently, the CC ensemble does not provide a
method suitable for the location of the coexistence point that has to be
estimated using the spontaneous creation method \cite{ezer}. It is expected 
that the difference observed between the models may be due to the different
behavior of the interface of the massive $CO$ cluster. 
So, we are planning to perform extensive simulations on 
this subject to clarify this open question.

\begin{figure}
\centerline{{\epsfysize=3.0 in \epsffile{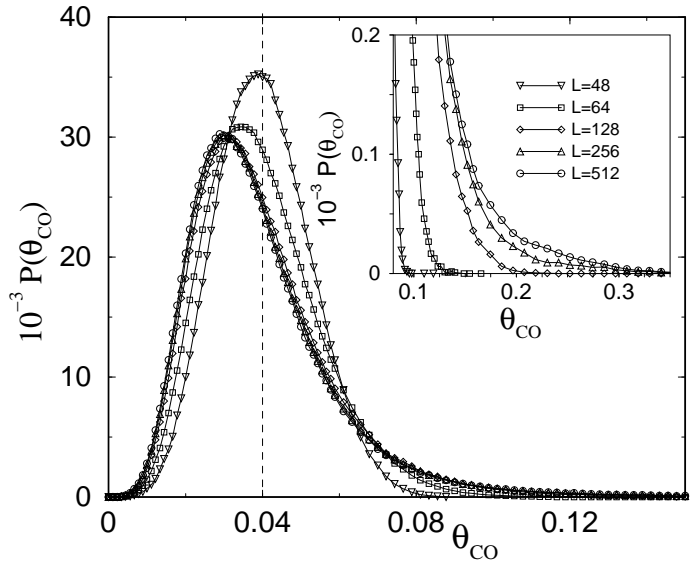}}}
\caption{ Plots of the {\it local} $CO-$coverage fluctuations 
$(P(\theta_{CO}))$ versus $\theta_{CO}$ measured using the constant 
coverage ensemble keeping the {\it global} $CO$ coverage at 
$\theta_{CO} = 0.040$ and using lattices
of different sizes as indicated in the figure, where $L$
is measured in LU. 
The inset shows a detailed view of the right-hand side of the 
distributions. Results averaged over $3 \times 10^6$
different configurations.}
\label{fig8}
\end{figure}

Pointing our attention to the upper spinodal point, it is found 
that its location depends on the lattice size, so $P^{US}_{CO}(L)$ 
can be determined from the loop, as 
shown in figures (2-4). It is expected that this dependence 
of $P^{US}_{CO}(L)$ is due to local fluctuations of the $CO$ coverage 
that take place during the
nucleation of the critical cluster. In order to check this conjecture,
the probability distribution of the {\it local} 
$CO$ coverage $(P(\theta_{CO}))$ in small
patches of side $L_{o} = 30 LU$ was measured for lattices 
of different sizes. It should be noticed that, in order to perform 
these measurements, the 
CC algorithm that keeps the {\it global} coverage
of $CO$ almost constant has been used. Setting $\theta_{CO} = 0.040$, i.e.
a value slightly smaller than the coverage at the upper spinodal point,
the probability distributions shown in figure 8 have been obtained.
For a rather small lattice ($L = 48$ in figure 8)  $P(\theta_{CO})$ 
is almost symmetric around the 
maximum $\theta_{CO}^{max} \approx \theta_{CO} = 0.040$. 
However, on increasing the lattice size the peak is shifted 
toward lower $\theta_{CO}$ values and the distribution function 
is clearly asymmetric. While
the left-hand side remains practically independent of finite size effects
(say for $L \geq 128 $ LU), an increasingly long tail emerges on the
right hand side of the distribution (for a detailed view see the inset of
figure 8). The existence of these non-vanishing tails implies
that fluctuations of the {\it local} $CO$ coverage up to relatively 
large values are present in larger samples. Such excursions of the
coverage induce the nucleation of critical $CO$ clusters in larger lattices
for smaller adsorption probabilities of $CO$, consequently $P^{US}_{CO}(L)$
must decrease upon increasing $L$.  
In fact, figure 9, which shows a plot of $P^{US}_{CO}(L)$
versus $L^{-1}$, confirms this trend. As a first approximation the 
data can be fitted by a straight
line that gives $P^{US}_{CO}(L\rightarrow\infty)\cong\,0.3544(2)$. 
Furthermore our estimate for the coverage 
is $\theta_{CO}^{US}\cong\,0.043(1)$. 
These results point out that in the thermodynamic limit the
spinodal point is very close to coexistence, i.e.
$\Delta\,P_{CO}=P_{CO}^{US}-P_{CO}^{Coex}\,\cong\,0.003$. For the sake of
comparison it is worth mentioning that for the ZGB model one has   
$\Delta\,P_{CO}\,\cong\,0.0012$ \cite{ezer}.
\begin{figure}
\centerline{{\epsfysize=3.0 in \epsffile{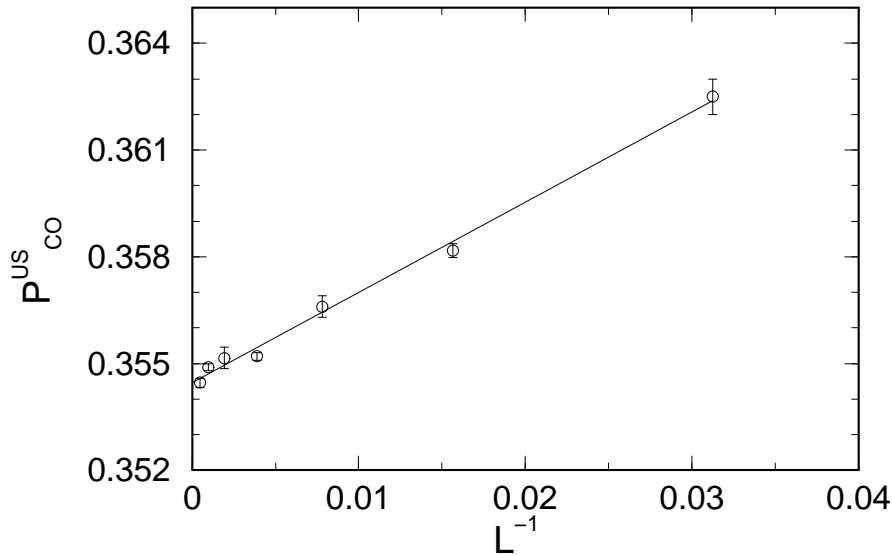}}}
\caption{ 
Plots of $P_{CO}^{US}$ versus $L^{-1}$, where $L$ is measured in LU. 
The straight line corresponds to the best fit of the data 
that extrapolates to $P_{CO}^{US}(L\rightarrow\infty)=0.3544(2)$.}
\label{fig9 }
\end{figure}

\subsection{Epidemic Study.}
In order to perform epidemic studies it is necessary to account 
for the fact that the poisoned (absorbing) state above coexistence 
is non-unique, since it is due to a mixture
of $CO$ and $N$ atoms with coverage $\theta_{CO}\,\approx\,0.9$ and
$\theta_{N}\,\approx\,0.1$, as shown in figure 1. So, the starting
configuration has to be obtained running the actual dynamics of the 
system slightly above coexistence until 'natural' absorbing states 
suitable for the studies are generated.

Extensive epidemic simulations ( $\sim 2$x$10^9$ different runs) 
have been performed for the following values of $P_{CO}$: $P_{CO}^{Coex}$,
$P_{CO}^{US}$, $P_{CO}^{DB}(L=256 LU)$ and $P_{CO}^{GB}(L=256 LU)$. A 
value close to coexistence but slightly inside the active region, 
namely $P_{CO}=0.347$, has also been used.
The obtained results can be observed in figure 10, which shows 
log-log plots of $N(t)$ versus $t$. It becomes evident that the 
method is quite sensitive to tiny changes of $P_{CO}$. The 
obtained curves are fitted by equation (\ref{anz}), as shown in 
figure 10 for $P_{CO}^{Coex}$. The best fits are obtained
for the parameters listed in Table I.

\begin{figure}
\centerline{{\epsfysize=3.0 in \epsffile{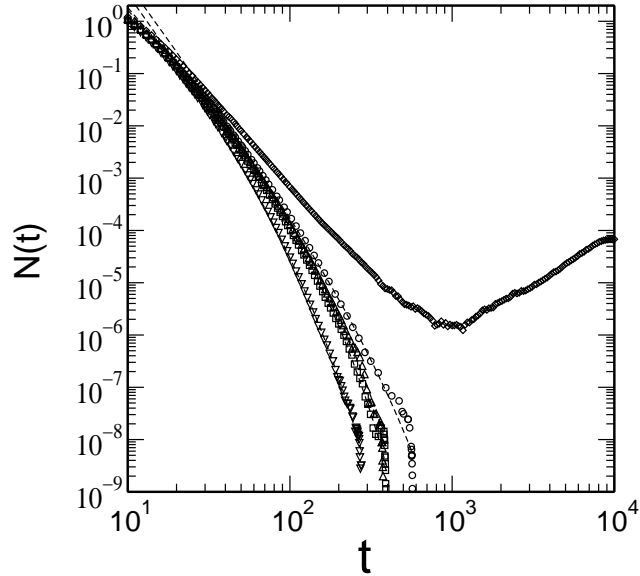}}}
\caption{ Log-log plots of the number of vacant sites $N(t)$ 
versus $t$; measured in MCS; for epidemic simulations performed 
using lattices of size $L=256$ LU. Results averaged up to $3$x$10^9$ different 
runs (triangle down) $P^{US}_{CO}=0.3544$, ($\square$) 
$P^{GB}_{CO}=0.3522$, ($\triangle$)$P^{Coex}_{CO}=0.35145$, ($\bigcirc$) 
$P^{DB}_{CO}=0.3506$  and ($\lozenge$)
 $P_{CO}=0.3470$)
.}
\label{fig10}
\end{figure}

It is concluded that close to coexistence one has 
$\eta_{eff}\,\cong\,3.5\pm\,0.5$ with a characteristic time $T \sim 50$ 
MCS except for $P_{CO}=0.3506$ with $T \sim 160$ MCS, a result that 
may reflect the fact that this point actually lies within the 
active region of the phase diagram. In fact, in the active region
larger $T$ values are expected, as judged by the results shown in figure 10.  

Of course, one may obtain better fits using different scaling 
ansatz than that in equation (\ref{anz}). Our main finding, however,  is 
that the occurrence of a power-law scaling behavior close to 
coexistence can unambiguously be ruled out. This
result is in qualitative agreement with recent numerical data obtained with
the ZGB model \cite{ezer}. All these observations are also in 
agreement with the experience gained studying first-order reversible 
phase transitions where it is well established that correlations 
are short ranged, preventing the emergency of scale invariance.

Let us now point our attention to the run performed taking 
$P_{CO}=0.347$ (figure 10). A rapid decrease in
$N(t)$ up to $10^3$ MCS due to the low survivability 
of the initial empty patch is observed. However, for $t>10^3$ MCS only 
few epidemics survive (actually around $10$ epidemics over $10^9$) 
and the average number of empty sites increases according 
to $N(t)\,\propto\,t^2$, indicating the homogeneous propagation 
of the epidemic. For $t\sim 10^4$ MCS the active region 
covers the whole sample and the average density of empty 
sites remains stationary close to $\theta_{emp}\,\approx\,0.52$.

\begin{center}
\vspace{0.5cm}

\begin{tabular}{|c|c|c|c|}
\hline
 $P_{CO}$ &
 $\eta _{eff}$ &
$T$ MCS &
 Number of epidemics \\ \hline

 $P^{GB}_{CO}=0.3506$ & $4.2\pm 0.2$ & $159\pm 10$ & $4.3$x$10^{9}$ \\
 \hline  $P^{Coex}_{CO}=0.35145$ &
 $3.45\pm 0.05 $ &
 $58\pm 1$ &
 $2.5$x$10^{9}$ \\
\hline
 $P^{DB}_{CO}=0.3522$ &
 $3.38\pm 0.01$ &
 $50\pm 1$ &
 $3.5$x$10^{9}$ \\ \hline
 $P^{US}_{CO}=0.3544$ & $4.1\pm 0.1$ & $40\pm 4$ & $5.5$x$10^{9}$ \\ 
\hline
\end{tabular}
\vspace{0.3cm}
\end{center}

{\bf Table I Caption.} Results obtained fitting the data corresponding to
different values of $P_{CO}$, shown in figure 10, using equation (5).

\section{Conclusions and outlook.}

In summary, after an extensive Monte Carlo simulation study 
of the first-order irreversible phase transition of the YK model
for the catalized $NO+CO$ reaction we concluded that: i) hysteretic effects are
found around the coexistence using the constant-coverage ensemble. 
The hysteresis loop basically exhibits three characteristic regions: a 
growing branch (GB), a vertical region (VR) and a decreasing branch (DB). 
Within each region the massive $CO$ cluster corresponding to the inactive 
phase has a well defined convex (GB), concave (DB) and
flat (VR) curvature. We are planning to study the dynamics of such interface 
in order to clarify the interplay between hysteresis 
and interfacial properties such as effective surface tension, 
roughness, curvature, etc. 
ii) The VR of the loop can be identified as the coexistence point. 
iii) The CC loop also gives evidence of a lattice size dependent 
upper spinodal point that can be extrapolated to the thermodynamic 
limit. Recently \cite{ezequi}  one of us has shown that 
the spinodal point can be located quite accurately performing 
studies of the short-time dynamics of the system.
So, we are planning to perform similar studies using the YK model in order to 
obtain an independent measurement of the upper spinodal point.
iv) Epidemic studies reveal the existence of short ranged correlations close
to coexistence.

Based on these findings and the experience gained studying a similar system 
(namely the ZGB model \cite{ezer}), we concluded that first-order 
irreversible phase transitions share many characteristics with 
their equilibrium (reversible) counterpart. So, we expect that these 
results will contribute to the development of a theoretical frame 
for the description of irreversible critical behavior.

\vskip 2.0 true cm

ACKNOWLEGMENTS: This work was supported by CONICET, UNLP, 
and ANPCyT (ARGENTINA). E. L. acknowledges the CIC for the provision of
a study fellowship.

\end{document}